\documentclass[useAMS,usenatbib]{mn2e}

\usepackage{amssymb,amsmath,upgreek,bm,color}

\usepackage{epsfig,graphicx,times,subfigure}
\usepackage[colorlinks=true, citecolor=cyan,
 linkcolor=cyan, menucolor=blue, urlcolor=blue,
 linkbordercolor={0 0 1}, frenchlinks=True, breaklinks]{hyperref}
\usepackage{orcidlink}

\newcommand{\oiii}{\hbox{[O\,{\sc iii}]}}
\newcommand{\ha}{\hbox{H$\alpha$}}
\newcommand{\hb}{\hbox{H$\beta$}}

\newcommand{\nii}{\hbox{[N\,{\sc ii}]}}

\newcommand{\ca}{\hbox{Ca\,{\sc ii}}}
\newcommand{\mg}{\hbox{Mg\,{\sc i}}}

\usepackage{gensymb}
\usepackage{mathrsfs}

\usepackage{url}

\title[Formation of CRDs in SDSS~J0748+4441]
{Uncovering the formation of the counter-rotating stellar disks in SDSS~J074834.64+444117.8}

\author[Bao, M. et al.]{
\parbox[t]{\textwidth}{\raggedright
Min~Bao~\orcidlink{0009-0005-9342-9125}$^{1,2,3}$,
Yanmei~Chen~\orcidlink{0000-0003-3226-031X}$^{1,2,3}$\thanks{E-mail: chenym@nju.edu.cn},
Meng~Yang~\orcidlink{0000-0002-1749-1892}$^{4}$\thanks{E-mail: myang@shao.ac.cn},
Ling~Zhu~\orcidlink{0000-0002-8005-0870}$^{4}$,
Yong~Shi~\orcidlink{0000-0002-8614-6275}$^{1,2,3}$,
Qiusheng~Gu~\orcidlink{0000-0002-3890-3729}$^{1,2,3}$
}\\
\vspace*{6pt}\\
$^{1}$Department of Astronomy, Nanjing University, Nanjing 210093, China\\ 
$^{2}$Key Laboratory of Modern Astronomy and  Astrophysics (Nanjing University), Ministry of Education, Nanjing 210093, China\\
$^{3}$Collaborative Innovation Center of Modern Astronomy and Space Exploration, Nanjing 210093, China\\
$^{4}$Shanghai Astronomical Observatory, Chinese Academy of Sciences, 80 Nandan Road, Shanghai 200030, China
	}

\begin{document}

\pagerange{\pageref{firstpage}--\pageref{lastpage}} \pubyear{}

\maketitle
\newpage
\label{firstpage}
\newpage
\pagebreak
\begin{abstract}
Using the integral field spectroscopic data from Mapping Nearby Galaxies at Apache Point Observatory survey, we study the kinematics and stellar population properties of the two counter-rotating stellar disks in a nearby galaxy SDSS~J074834.64+444117.8. We disentangle the two stellar disks by three methods, including {\ca}~$\lambda$8542 double Gaussian fit, pPXF spectral decomposition, and orbit-based dynamical model. These three different methods give consistent stellar kinematics. The pPXF spectral decomposition provides the spectra of two stellar disks, with one being more luminous across the whole galaxy named primary disk, and the other named secondary disk. The primary disk is counter-rotating with ionized gas, while the secondary disk is co-rotating with ionized gas. The secondary disk has younger stellar population and poorer stellar metallicity than the primary disk. We estimate the stellar mass ratio between the primary and secondary disks to be $\sim$5.2. The DESI $g$, $r$, $z$ color image doesn't show any merger remnant feature in this galaxy. These findings support a scenario that the counter-rotating stellar disks in SDSS~J074834.64+444117.8 formed through gas accretion from the cosmic web or a gas-rich companion.
\end{abstract}


\section{Introduction}
\label{sec: introduction}

% Definition
Counter-rotating stellar disks are characterized by the presence of two stellar disks that are co-spatial in one galaxy but rotating in opposite directions. Such a feature was first discovered in NGC~4550 by absorption line analyses \citep{1992ApJ...394L...9R, 1992ApJ...400L...5R}. Moreover, 2$\sigma$ feature was proposed by \cite{2011MNRAS.414.2923K} to describe the observation of off-center but symmetric peaks along the major axis in the stellar velocity dispersion field of a galaxy with counter-rotating stellar disks, where the co-existing blueshifted and redshifted absorption components contribute to the broadening of absorption lines.

% Individual studies (spectral decomposition: external gas fuels the disk formation)
Individual galaxies hosting counter-rotating stellar disks have been studied, 
including NGC~5719 \citep{2011MNRAS.412L.113C}, NGC~4550 \citep{2013MNRAS.428.1296J}, NGC~3593 \citep{2013A&A...549A...3C}, NGC~4138 \citep{2014A&A...570A..79P}, NGC~448 \citep{2016MNRAS.461.2068K}, NGC~5102 \citep{2017MNRAS.464.4789M}, and IC~719 \citep{2018A&A...616A..22P}. Using the spectral decomposition method, it is possible to disentangle the properties of the co-spatial stellar disks. \cite{2013MNRAS.428.1296J} decomposed the spectra along the major axis of S0 galaxy NGC~4550, finding that the stellar disk that co-rotates with the gas disk has younger stellar population than the counter-rotating one, which indicates the co-rotating stellar disk lately forming from the accreted gas. \cite{2011MNRAS.412L.113C} applied the decomposition method to the spectra of spiral galaxy NGC~5719, and revealed that the co-rotating stellar disk not only has younger stellar population, but also has poorer stellar metallicity. Considering that NGC~5719 is interacting with a companion NGC~5713, \cite{2011MNRAS.412L.113C} suggested these findings as the accreted gas fueling the in-situ formation of a new stellar disk.

% Sample studies (spectral decomposition: )
With the development of integral field spectroscopic (IFS) surveys, studies on samples of counter-rotating stellar disks became feasible, which revealed the dependence of formation scenarios on numerous physical factors. \cite{2022MNRAS.511..139B} selected 64 galaxies with counter-rotating stellar disks from $\sim$4,000 galaxies in the MaNGA survey Data Release~16, and found that in most cases the younger stellar disk co-rotates with the gas disk, which supported the formation scenario of gas accretion. However, the gas co-rotates with the older stellar disk in two galaxies, which implies a disk galaxy merging with a gas-poor galaxy with younger stellar population in a retrograde orbit. Moreover, \cite{2022ApJ...926L..13B} collected a sample of 101 galaxies with counter-rotating stellar disks from 9,456 galaxies in the MaNGA survey Product Launch~10, divided the sample into four types based on the stellar and gas kinematics, and proposed different formation scenarios for different types. They suggested that the key factors in the formation of counter-rotating stellar disks are the abundance of pre-existing gas in the progenitor and the efficiency of angular momentum consumption between the pre-existing and external gas. Given the complexity of formation mechanisms, it is necessary to adopt reliable methods to disentangle two stellar disks, and compare properties between them as completely as possible to uncover the formation of counter-rotating stellar disks.

% Orbit-based model
Orbit-based dynamical model provides a physical method to quantify the structure and dynamics of different galaxy components based on their orbit structure distribution. This method has been applied to galaxies to explore the relation between galaxy components and galaxy assembly history in several IFS survey, such as CALIFA~\citep{2018MNRAS.479..945Z}, MaNGA~\citep{2020MNRAS.491.1690J}, SAMI~\citep{2022ApJ...930..153S} and MUSE~\citep{2023A&A...672A..84D}. And the method has recently been updated to include bar properly~\citep{2022ApJ...941..109T}. However, it has not been applied to the galaxies with counter-rotating stellar disks, since the line-of-sight velocity distribution (LOSVD) of these galaxies cannot be well described by the widely-used Gaussian-Hermite expansion of single dynamical component~\citep{2021A&A...654A..30R}. The solution to this problem relies on the non-parametric LOSVDs which capture the multiple dynamical components~\citep[e.g.][]{2011BaltA..20..453K, 2021A&A...646A..31F}, providing a new prospect of modelling counter-rotating stellar disks with orbit-based dynamical models and public DYNAMITE package~\citep{2008MNRAS.385..647V, 2020ascl.soft11007J, 2022A&A...667A..51T}.

% Paper structure
In this paper, we focus on a nearby ($z\sim0.02$) galaxy SDSS~J074834.64+444117.8 (hereafter SDSS~J0748+4441). We disentangle the two stellar disks by three methods, including {\ca}~$\lambda$8542 double Gaussian fit, pPXF spectral decomposition, and orbit-based dynamical model. The involved data are presented in Section \ref{sec: data}. The spatially resolved galaxy parameters, the disentangle results, as well as the properties of the two stellar disks are presented in Section \ref{sec: results}. Finally, we discuss the formation scenario for counter-rotating stellar disks in Section \ref{sec: discussion}.

\section{The Data}
\label{sec: data}

% MaNGA data
MaNGA is an integral field spectroscopic (IFS) survey conducted as a part of the fourth-generation Sloan Digital Sky Survey (SDSS-IV; \citealt{2015ApJ...798....7B, 2016AJ....152...83L}). MaNGA survey provides a representative sample of 10,010 unique galaxies with a flat stellar mass distribution in the range of $10^{9}-10^{11}\rm M_{\odot}$ and redshift in the range of $0.01<z<0.15$ \citep{2017AJ....154...28B}. This survey utilizes the Baryon Oscillation Spectroscopic Survey (BOSS) spectrographs \citep{2013AJ....146...32S} on the 2.5-m Sloan Foundation Telescope \citep{2006AJ....131.2332G}. The dual-channel BOSS spectrographs \citep{2013AJ....146...32S} provide simultaneous wavelength coverage from 3,600 to 10,000~\AA.

% DRP & DAP
For each target, the MaNGA Data Reduction Pipeline (DRP; \citealt{2016AJ....152...83L}) produced sky-subtracted spectrophotometrically calibrated spectra, and generated three-dimensional datacube containing spatially resolved spectra. The wavelength calibration of the MaNGA data is accurate to 5~km~s$^{-1}$~(rms), with a median spectral resolution of 72~km~s$^{-1}$ ($R\sim$2,000). Moreover, the MaNGA Data Analysis Pipeline (DAP; \citealt{2019AJ....158..231W}) provided measurements with pixel size of 0.5$^{\prime\prime}$/spaxel ($\sim$211~pc/spaxel), including stellar kinematics, ionized gas kinematics, emission line flux and equivalent width, as well as spectral indices such as 4000~\AA~break indicating the light-weighted stellar population age. The gas kinematics in this paper is traced by {\ha} emission, while all the emission line centers are tied together in the velocity space in MaNGA DAP. Besides, we extract the stellar mass in each spaxel from \texttt{Pipe3D} \citep{2016RMxAA..52...21S}, which used a spectral fitting tool \texttt{FIT3D} to analyse the physical properties of stellar populations of a galaxy.

% SDSS~J074834.64+444117.8
SDSS~J0748+4441 was chosen from a sample of 101 galaxies hosting counter-rotating stellar disks (CRDs; \citealt{2022ApJ...926L..13B}) since two distinguishable absorption components are clearly shown in one of the {\ca} triplet lines at $\lambda$8542 which is free from skylines. The MaNGA spectra have high enough signal-to-noise ratio (SNR) for disentangling the absorptions of two stellar disks even on the outskirts ($\sim$1.5~$Re$). The SDSS $g$, $r$, $i$ color image of this galaxy is displayed in Figure \ref{fig:kinematics}(a). The global stellar mass ($M_{\star}$) and attenuation corrected star formation rate (SFR) from the MaNGA DRP and DAP indicate that SDSS~J0748+4441 locates at the green-valley region.

\section{The Results}
\label{sec: results}

\subsection{Spatially resolved properties}

Figure \ref{fig:kinematics} displays the kinematics of stars and ionized gas in SDSS~J0748+4441. Figures \ref{fig:kinematics}(b) and \ref{fig:kinematics}(c) display the stellar velocity and velocity dispersion fields for spaxels with median spectral SNR per spaxel higher than 3, where the black dashed line shows the photometric major axis (hereafter major axis). The two black stars in Figure \ref{fig:kinematics}(c) mark the 2$\sigma$ peaks along the major axis, where the maximum stellar velocity dispersion locates. The presence of two regions with enhanced stellar velocity dispersion along the major axis originates from the existence of two CRDs. Meanwhile, the stellar velocity field in Figure \ref{fig:kinematics}(b) following a regular pattern indicates that it is dominated by the more luminous stellar disk across the whole galaxy.

Figures \ref{fig:kinematics}(e) and \ref{fig:kinematics}(f) display the ionized gas velocity and velocity dispersion fields for spaxels with {\ha} emission line SNR higher than 3. The gas is regularly rotating as displayed in Figure \ref{fig:kinematics}(e), with low gas velocity dispersion around the 2$\sigma$ peaks (black stars) in Figure \ref{fig:kinematics}(f). We fit the gas velocity field using the \texttt{KINEMETRY} package \citep{2006MNRAS.366..787K} to derive the kinematic major axis, which is shown by the grey dashed lines in Figures \ref{fig:kinematics}(e) and \ref{fig:kinematics}(f). Combining the stellar and gas velocity fields in Figures \ref{fig:kinematics}(b) and \ref{fig:kinematics}(e), we conclude that the more luminous stellar disk and the gas disk in SDSS~J0748+4441 are counter-rotating. Referring to the kinematic classification in \cite{2022ApJ...926L..13B}, this galaxy belongs to Type~2b, where the old stellar disk outshines the newly formed one and counter-rotates with gas disk. In Section \ref{sec: disentangling}, we will disentangle and compare the properties of two stellar disks.

\begin{figure*}
    \resizebox{0.8\textwidth}{!}{\includegraphics{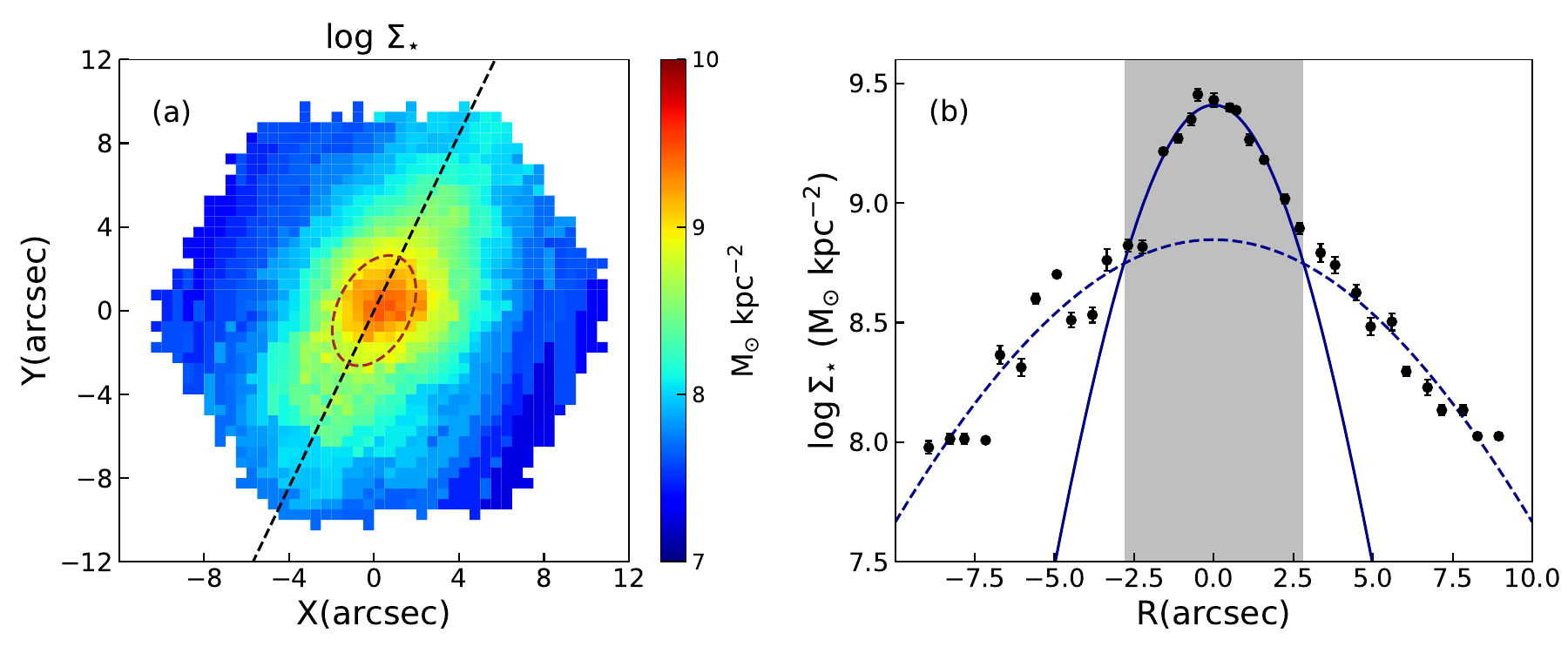}}
    \caption{Spatially resolved stellar mass surface density. (a) A map of stellar mass surface density for spaxels with median spectral SNR per spaxel higher than 3. The red dashed ellipse with a half major axis of $|R|=2.8^{\prime\prime}$ outlines the bulge region in Figures \ref{fig:mass} and \ref{fig:emission}. (b) The black circles represent the stellar mass surface density at different radii along the major axis. The black vertical bar shows the stellar mass surface density $\pm1\sigma$ error at each radius. The grey area marks the bulge region with $|R|\leq2.8^{\prime\prime}$ in Figures \ref{fig:mass}, \ref{fig:rotation_curve}, \ref{fig:stellar_properties} and \ref{fig:disks_ratio}. The blue solid profile shows the gaussian model for the data points inside the grey area. The blue dashed profile shows the gaussian model for the data points outside the grey area.}
    \label{fig:mass}
\end{figure*}

Figure \ref{fig:mass}(a) displays a map of stellar mass surface density ($\Sigma_{\star}$), for spaxels with median spectral SNR per spaxel higher than 3. The $\Sigma_{\star}$ is defined as the stellar mass of each spaxel divided by the physical size of the spaxel. Figure \ref{fig:mass}(b) displays $\Sigma_{\star}$ as a function of radius along major axis. $\Sigma_{\star}$ monotonically decreases with increasing radius, but the gradients in the central region and on the outskirts are obviously different. The grey area with $|R|\leq2.8^{\prime\prime}$ separates the central region with high surface density from the outskirts with low surface density. We fit two Gaussian functions to the data points inside and outside the grey area. The blue solid profile shows the Gaussian model for the $\Sigma_{\star}$ gradient with $|R|\leq2.8^{\prime\prime}$, while the blue dashed profile shows the Gaussian model for the data points with $|R| > 2.8^{\prime\prime}$. It is clear that the central region and outskirts follow different $\Sigma_{\star}$ distributions, dominated by bulge and disk, respectively. We mark a red dashed ellipse with a half major axis of $|R|=2.8^{\prime\prime}$ on the $\Sigma_{\star}$ map, which basically covers the densest bulge region.

\begin{figure*}
\resizebox{0.8\textwidth}{!}{\includegraphics{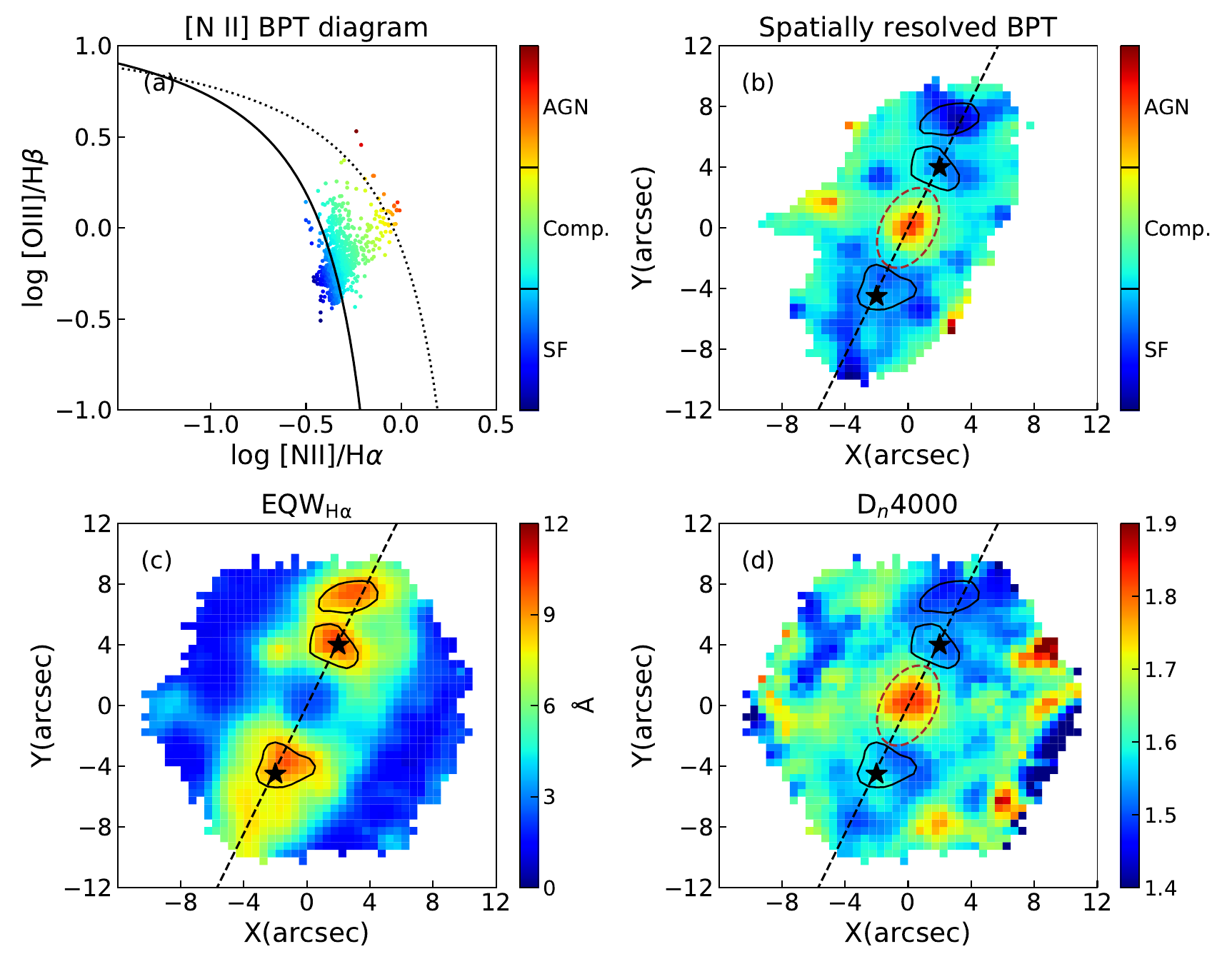}}
\caption{BPT diagrams and spatially resolved properties. (a) {\nii} BPT diagram. All the spaxels satisfy the H$\beta$, {\oiii}$\lambda$5007, {\ha} and {\nii}$\lambda$6583 emission line SNRs higher than 3. The black solid curve shows the demarcation between the star forming and composite regions \citep{2003MNRAS.346.1055K}, while the black dotted curve shows the demarcation between the composite and AGN regions \citep{2001ApJ...556..121K}. The dots show the line ratios measured in different spaxels, and they are color-coded by the distances of corresponding spaxels to the black solid curve. (b) Spatially resolved BPT diagram. The SNR criteria and color-code are the same as panel (a). (c) A map of {\ha} equivalent width for spaxels with {\ha} emission line SNR higher than 3. The black contours outline the regions with enhanced {\ha} EQW in panels (c) and (d). (d) A map of 4000~\AA~break (D$_{n}$4000) with median spectral SNR per spaxel higher than 3.}
\label{fig:emission}
\end{figure*}

Mapping diagnostic emission line ratios across a galaxy gives the information on the ionization state distribution of gas. Figure \ref{fig:emission}(a) displays the {\nii} BPT diagram \citep{1981PASP...93....5B} for spaxels with H$\beta$, {\oiii}$\lambda$5007, {\ha}, and {\nii}$\lambda$6583 emission line SNRs higher than 3. The black solid curve separates the star forming and composite regions \citep{2003MNRAS.346.1055K}, while the black dashed curve is the demarcation between the composite and AGN regions \citep{2001ApJ...556..121K}. The dots show the line ratios measured in different spaxels, and they are color-coded by the distances of corresponding spaxels to the black solid curve. Figure \ref{fig:emission}(b) displays the spatially resolved BPT diagram, where the color-codes are the same as Figure \ref{fig:emission}(a). The over-plotted red dashed ellipse is the same as that in Figure \ref{fig:mass}(a) with a half major axis of $|R|=2.8^{\prime\prime}$. Regions inside the ellipse are dominated by AGN, while the outer regions including the 2$\sigma$ peaks (black stars) are dominated by star formation.

The star formation rate describes the ongoing activity of star formation that occurred within the last 10$^{6-7}$~years. We derive the global SFR of SDSS~J0748+4441, which is $\sim$6.7$\times$10$^{-2}$~M$_{\odot}\rm~yr^{-1}$, based on attenuation corrected {\ha} flux from MaNGA DAP. Figure \ref{fig:emission}(c) displays a map of {\ha} equivalent width (EQW), which is a good indicator of the specific star formation rate in the star forming region. The black contours in Figure \ref{fig:emission}(c) outline three regions with enhanced {\ha} EQW, which are also marked in Figure \ref{fig:emission}(b). It is clear that the {\ha} EQW enhanced regions are dominated by star formation, indicating a higher sSFR in these areas. It is interesting that the two black stars marking the 2$\sigma$ peaks locate within the contours with the enhanced {\ha} EQW.

Figure \ref{fig:emission}(d) displays a map of 4000~\AA~break (D$_{n}$4000), which is influenced by the intensity of star formation in Gyr-timescale. Along the major axis, D$_{n}$4000 presents the highest value of $\sim$1.8 in the bulge region (inside the red dashed ellipse), while it has a value of $\sim$1.5 in the disk region, including the 2$\sigma$ peaks (black stars) and the enhanced {\ha} EQW regions (black contours). The homologous D$_{n}$4000 distribution but clumpy {\ha} EQW enhancement in the disk region can be explained by the different star formation timescales represented by {\ha} emission and D$_{n}$4000.

\subsection{Disentangling two stellar disks}
\label{sec: disentangling}

\subsubsection{Double Gaussian fit on {\ca}~$\lambda$8542}

SDSS~J0748+4441 shows obvious asymmetric absorption structure in one of the {\ca} triplet lines at $\lambda$8542, which provides us an opportunity to disentangle the two stellar disks without model dependence. The other two of {\ca} triplet lines at $\lambda\lambda$8498, 8662 are contaminated by the skylines.

Figure \ref{fig:spectrum_fit}(a) displays the stellar velocity dispersion field given by MaNGA DAP, in which the two black rings outline the regions around the 2$\sigma$ peaks (black stars), with the top one named `Region A' and the bottom one named `Region B'. To intuitively display the two absorption components in the {\ca}~$\lambda$8542 absorption line, we extract and stack the spectra within Regions A and B, respectively. The {\ca}~$\lambda$8542 absorption lines in the stacked spectra of Regions A and B are shown in the Figures \ref{fig:spectrum_fit}(b) and \ref{fig:spectrum_fit}(c), where we use the python-based tool \texttt{curve\_fit} to conduct double Gaussian fit on them. The Gaussian models with stronger absorption are in red, the Gaussian models with weaker absorption are in blue, and the best-fit models are in green which is the combination of blue and red components.

\subsubsection{pPXF spectral decomposition}

Taking advantage of the different velocities and velocity dispersions of two CRDs, we decompose their contributions to the observed spectrum in each spaxel along the major axis, using the penalized pixel fitting (pPXF; \citealt{2017MNRAS.466..798C}) code. The code builds two synthetic templates (one for each stellar disk) as linear combination of stellar spectra from the MILES library, convolves the templates with two Gaussian LOSVDs with different velocities and velocity dispersions, and ensures the combination of two convolved templates fitted to the observed spectrum by $\chi^{2}$ minimization.

Figure \ref{fig:spectrum_fit}(d) displays an example of pPXF two kinematic component fit. The observed spectrum shown in black is extracted from the spaxel marked by the black filled star in Figure \ref{fig:spectrum_fit}(a). The red and blue spectra show the optimal models for the two stellar disks. The red spectrum corresponds to the stellar disk with higher flux (hereafter primary disk), and the blue spectrum corresponds to the stellar disk with lower flux (hereafter secondary disk). The green spectrum is the combination of the primary and secondary components, it is our best-fit model. The insert drawing of Figure \ref{fig:spectrum_fit}(d) highlights the spectra in [4800, 5300]~\AA~wavelength range. The four grey dashed lines mark the rest-frame wavelength of {\hb} and {\mg} triplet lines, where all the absorption profiles are well fitted.

\begin{figure*}
\resizebox{0.7\textwidth}{!}{\includegraphics{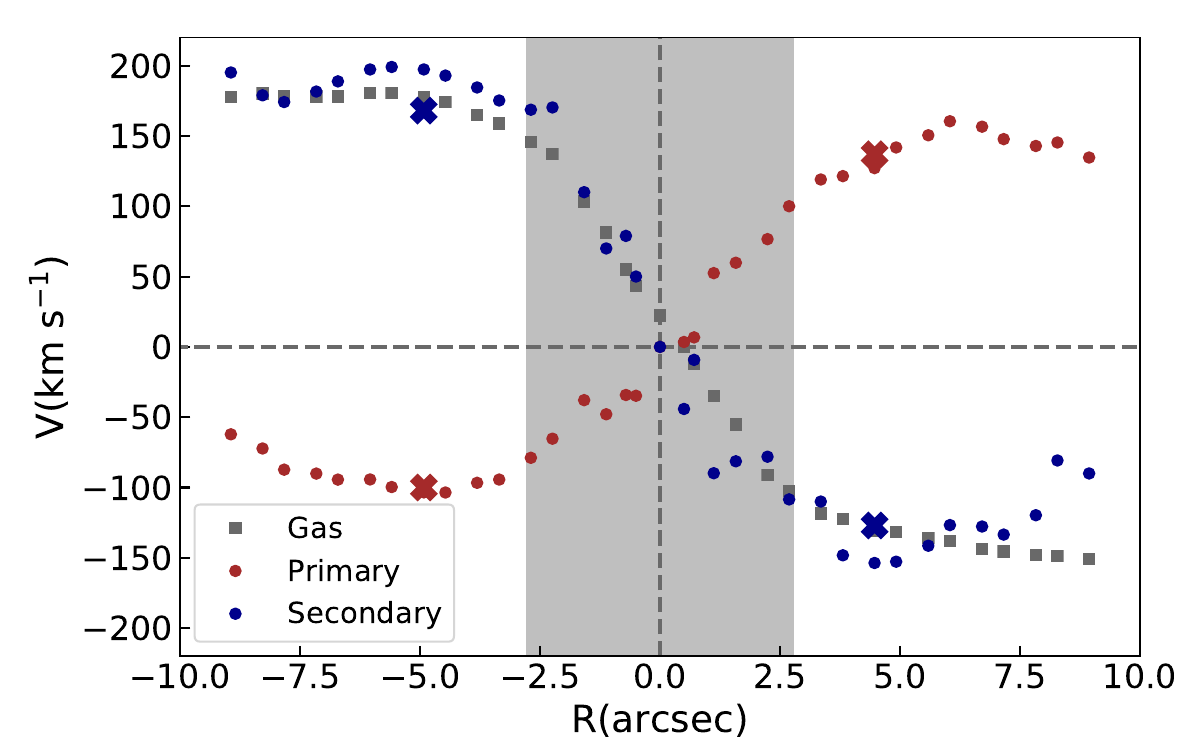}}
\caption{Line-of-sight velocities as functions of radii along the major axis. The circles represent the velocities from the pPXF spectral decomposition, with primary and secondary disks in red and blue. The red (primary disk) and blue (secondary disk) crosses represent the velocities from double Gaussian fit on {\ca}~$\lambda$8542. The grey squares represent the gas velocity traced by {\ha} emission from the MaNGA DAP file.}
\label{fig:rotation_curve}
\end{figure*}

Figure \ref{fig:rotation_curve} displays the line-of-sight velocities of stars and gas as functions of radii. The line-of-sight velocities of the two stellar disks are obtained by the pPXF two kinematic component fit on the continuum and absorption lines for spaxels along the major axis. The red and blue circles in Figure \ref{fig:rotation_curve} represent the line-of-sight velocities of the primary and secondary disks, respectively. The line-of-sight velocity of the primary disk is overall lower than that of the secondary disk. We also obtain stellar velocity dispersion by the pPXF two kinematic component fit. The stellar velocity dispersions of both primary and secondary disks follow flat distribution along the major axis, with comparable values of $84\rm~km~s^{-1}$ and $79\rm~km~s^{-1}$, respectively. The grey area marks the bulge region, where the stellar kinematics of primary disk is dominated by velocity dispersion, resulting in larger decomposition error. 

We collect the velocity of the ionized gas traced by {\ha} from the MaNGA DAP file \citep{2019AJ....158..231W}, which are represented by the grey squares in Figure \ref{fig:rotation_curve}. The line-of-sight velocities of the secondary disk (blue circles) and the gas disk (grey squares) are totally consistent, indicating the secondary disk is co-rotating with the gas disk. Meanwhile, the primary disk (red circles) is counter-rotating with the gas disk. The fit on {\ca}~$\lambda$8542 lines in Figures \ref{fig:spectrum_fit}(b) and \ref{fig:spectrum_fit}(c) also provides velocities of two stellar disks in Regions A and B, which are shown as the crosses in Figure \ref{fig:rotation_curve}, with the primary and secondary disks in red and blue, respectively. The line-of-sight velocities of the two stellar disks measured from {\ca}~$\lambda$8542 absorption lines match well with the results from the pPXF spectral decomposition.

\subsubsection{Orbit-based dynamical model}

The Schwarzschild orbit-superposition method \citep{1979ApJ...232..236S} is a powerful orbit-based dynamical modelling technique to reveal orbit structures in a galaxy. This method computes all possible stellar orbits in a given potential and assigns weight to each orbit to recover galaxy kinematics, including widely used parametric LOSVDs (e.g., Gaussian-Hermite expansion with velocity and velocity dispersion) in ordinary galaxies and non-parametric LOSVDs which are superior in describing the counter-rotating spectral features generated by two stellar disks. Compared with the spectral decomposition method, it makes use of spatial information contained in all IFS spectra of a galaxy and sidesteps the high SNR requirement when decomposing two stellar components directly from the spectra. In this section, we confirm the kinematics of the CRDs in SDSS~J0748+4441 with the orbit-based dynamical models using the non-parametric LOSVDs of this galaxy.

\begin{figure*}
  \resizebox{0.98\textwidth}{!}{\includegraphics{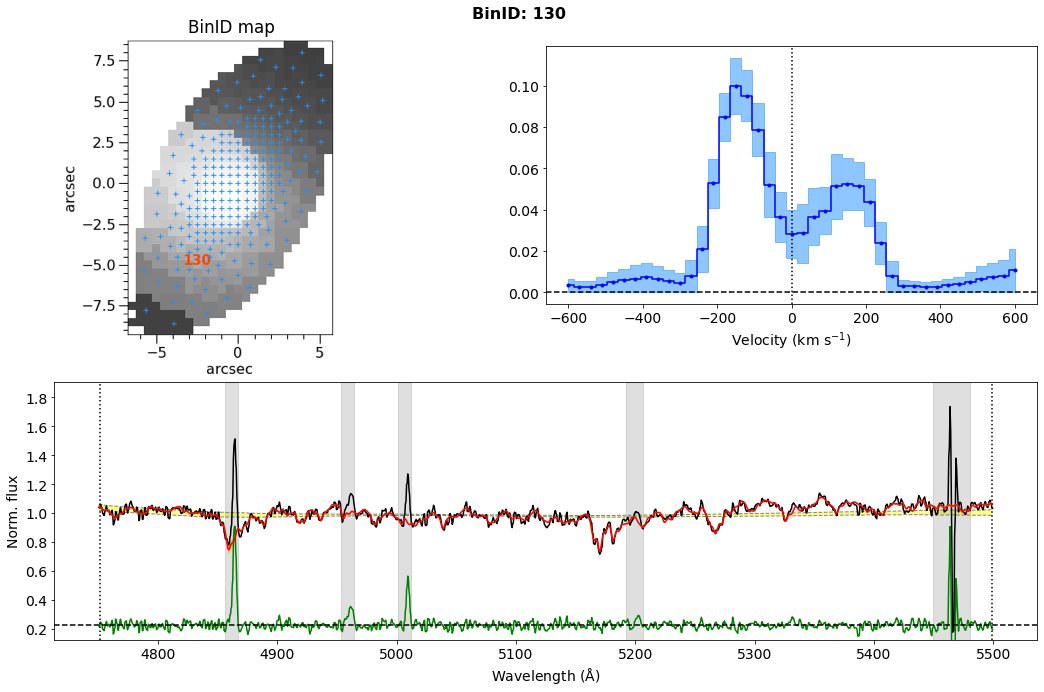}}
\caption{Spectra fit and LOSVD of bin 130 in orbit-based dynamical model generated by \texttt{BAYES-LOSVD}. Top left: the binning map with the position of bin 130 marked by the red id. Top right: the blue solid line and shadow are the LOSVD and $\pm1\sigma$ error extracted by fitting the MaNGA spectra with \texttt{BAYES-LOSVD}. Bottom: MaNGA spectra (black), fit (red) and residual (green) of bin 130. The emission lines covered by grey area are masked prior to fit.}
\label{fig:bin130}
\end{figure*}

We first extract the non-parametric LOSVDs of this galaxy from the MaNGA datacube with the \texttt{BAYES-LOSVD} package\footnote{\url{https://github.com/jfalconbarroso/BAYES-LOSVD}}, which employs Bayesian inference to obtain LOSVDs and the corresponding uncertainties. This package improves its efficiency by adopting a principal component analysis to reduce the requirement of stellar templates, and provides several prior options to regularize the output LOSVDs. We refer readers to \cite{2021A&A...646A..31F} for more details. For SDSS~J0748+4441, the MaNGA spectra are first Voronoi binned to SNR $>$ 50 using the median SNR within the fitting spectra range [4750, 5500]~\AA, and then fitted with the MILES stellar template library~\citep{2006MNRAS.371..703S,2011A&A...532A..95F}. The output LOSVDs have a velocity range [-600, 600]~km/s with a step of 30 km/s ($\sim$1/2 MaNGA instrumental resolution). As an example, we show the LOSVD extraction of bin~130 in Figure~\ref{fig:bin130}, which is the closest to the black-filled star in Figure \ref{fig:spectrum_fit}(a).

\begin{figure*}
\resizebox{0.98\textwidth}{!}{\includegraphics{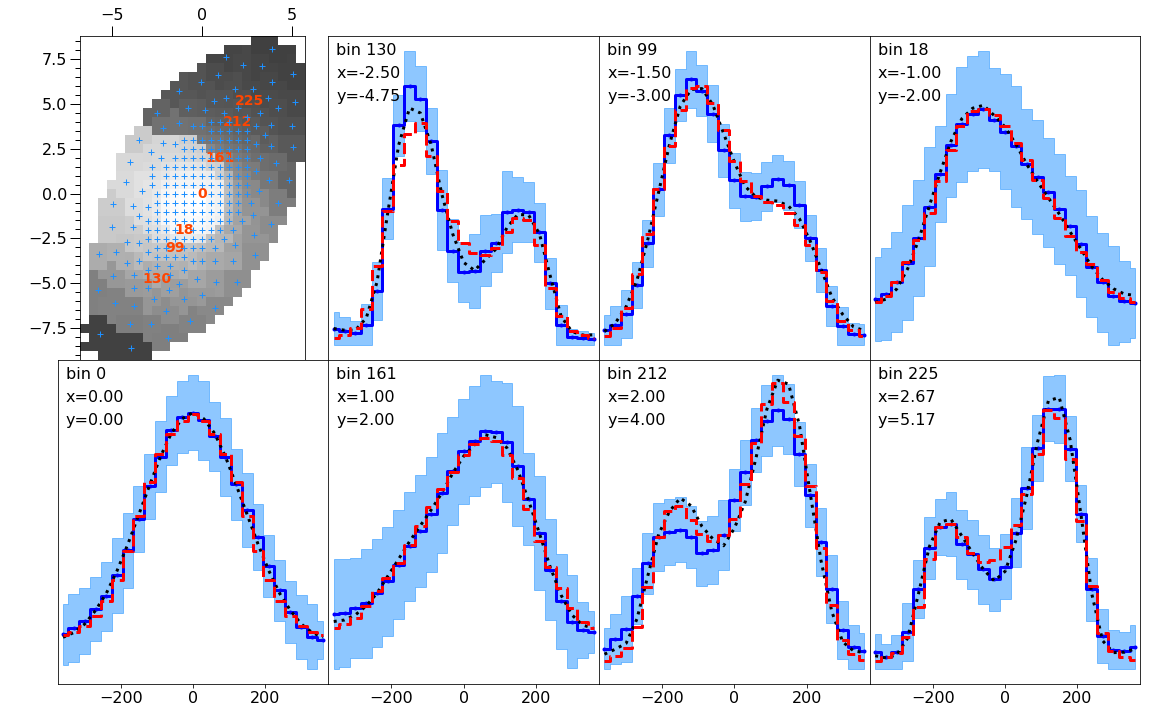}}
\caption{Fitting LOSVD along the major axis in orbit-based dynamical model. First panel: the binning map showing the ids and positions (in arcsec) of all the bins shown in the following panels. Other panels: the blue solid line and shadow are the LOSVD and $\pm1\sigma$ error obtained with \texttt{BAYES-LOSVD}, the red dashed line are the best-fit model, and the black dotted line are the symmetrized data for comparison because the model is axis-symmetric.}
\label{fig:major-fitting}
\end{figure*}

We then construct orbit-based dynamical models from the non-parametric LOSVDs using the \texttt{DYNAMITE} package~\citep{2008MNRAS.385..647V,2020ascl.soft11007J,2022A&A...667A..51T} and obtain the orbit distribution of this galaxy. We build the stellar mass distribution of this galaxy from its $r$-band image \citep{2002SPIE.4836..154K} with the MGE method~\citep{1994A&A...285..723E,2002MNRAS.333..400C}. Moreover, we correct for its mass-to-light ratio with the stellar mass obtained from the datacube of \texttt{Pipe3D}~\citep{2018RMxAA..54..217S}, and introduce a variable to indicate the scale of the total stellar mass. We adopt the NFW dark matter profile described with virial mass $M_{200}$ and concentration $c$. $M_{200}$ is defined as the enclosed mass within the virial radius $r_{200}$, where the average density is 200 times the critical density ($\rho_\mathrm{crit} = 1.37\times 10^{-7} \mathrm{M_\odot~pc^{-3}}$), and $c$ is defined as the ratio of $r_{200}$ to the scale radius of the NFW dark matter profile. We also include a central black hole with a fixed mass of $10^6~\rm M_\odot$. More details will be described in Yang et al. (in preparation). We show the fit along the major axis for SDSS~J0748+4441 in Figure~\ref{fig:major-fitting}, which includes bin~130 mentioned above. 

\begin{figure*}
    \resizebox{0.49\textwidth}{!}{\includegraphics{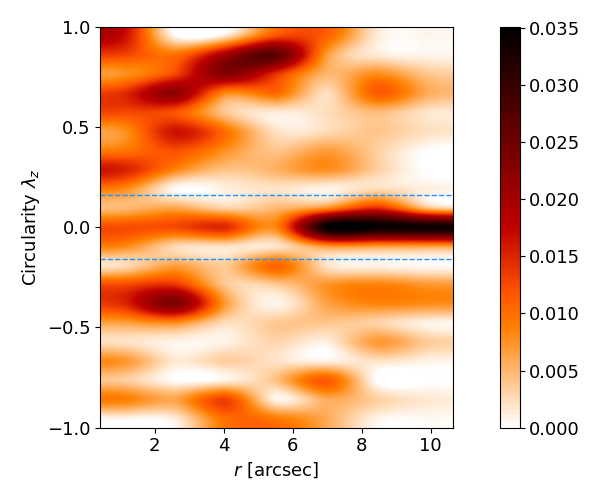}}
    \resizebox{0.49\textwidth}{!}{\includegraphics{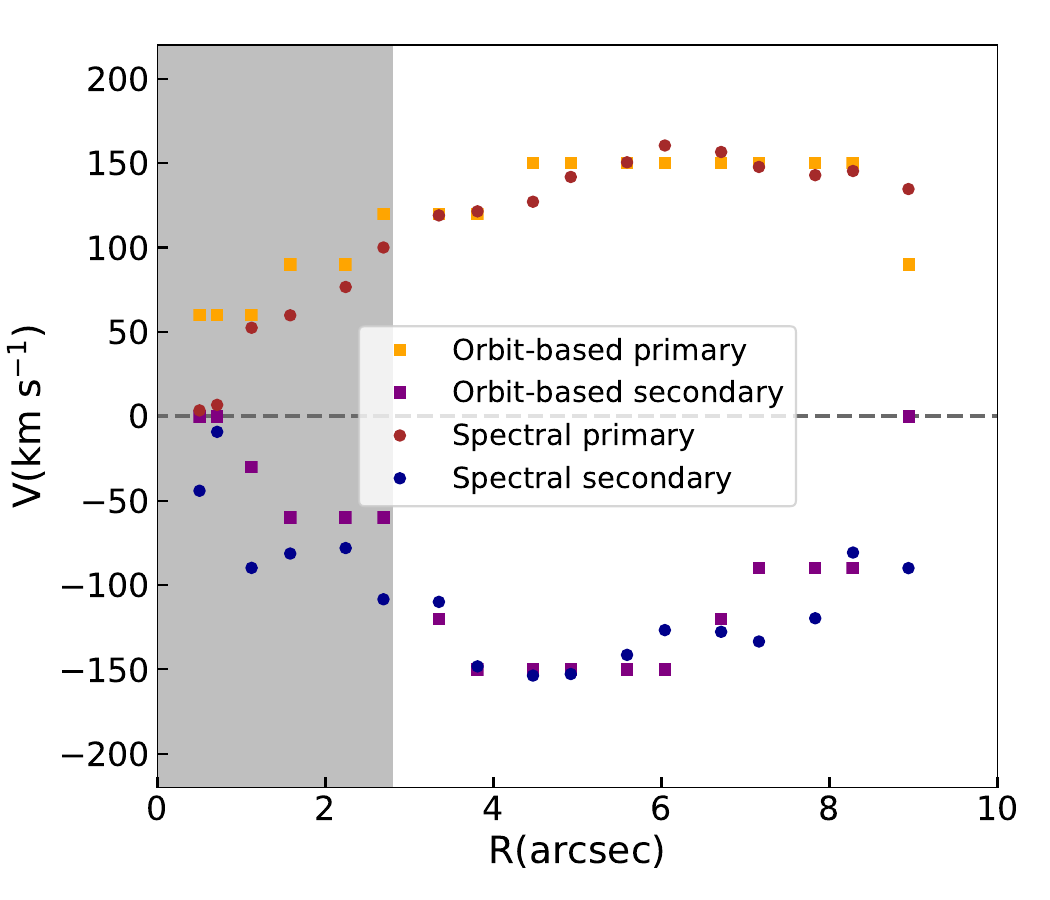}}
    \caption{Best-fitting orbit-based dynamical model on the phase-space and the resulting line-of-sight velocities. First panel: The probability density of stellar orbits on the phase-space of circularity $\lambda_z$ versus average radius $r$ of the best-fitting model. The blue dashed lines indicate $\lambda_z=\pm 0.16$. The colorbar shows the probability density of orbits. Second panel: The red and blue circles are the same as Figure \ref{fig:rotation_curve}. The squares represent the velocities from the orbit-based dynamical model, with primary and secondary disks in orange and purple.}
    \label{fig:circularity}
\end{figure*}

We further characterize different orbit types with the circularity
\begin{equation}
    \lambda_z = \overline{L_z}/r\overline{V_c}.
\end{equation}
For each orbit, $\overline{L_z}$ is the mean angular momentum along the short axis, $r$ is the average radius in equal time step for each orbit, and $\overline{V_c}$ is the mean circular velocity (see Equation 9 in \citealt{2018MNRAS.473.3000Z} for more detailed mathematical definitions). We divide the orbits into three stellar components by clear circularity distinctions at  $\lambda_z = \pm 0.16$, shown as two blue dashed lines in the first panel of Figure \ref{fig:circularity}. The secondary disk is comprised of the co-rotating component classified as $\lambda_z > 0.16$, while the primary disk is comprised of the non-rotating and counter-rotating components classified as $-0.16 < \lambda_z < 0.16$ and $\lambda_z < -0.16$. In the second panel of Figure \ref{fig:circularity}, we display the line-of-sight orbital velocities as functions of radii along the top half of the major axis for the two stellar disks, with the primary and secondary disks represented by the orange and purple squares. The line-of-sight velocities obtained by the orbit-based model are consistent with those obtained by the spectral decomposition, with the primary and secondary disks represented by red and blue circles in the second panel of Figure \ref{fig:circularity}, respectively.

\subsubsection{Properties of two stellar disks}

Based on the disentangled spectra of the two stellar disks in SDSS~J0748+4441, we compare the properties between them. We measure the Lick indices {\hb}, Mgb, Fe5270, and Fe5335 \citep{1994ApJS...94..687W}, for the primary and secondary components. {\hb} can be used as an indicator of stellar population age. We also calculate the combined magnesium-iron index $\rm [MgFe]^{\prime}$ as $\sqrt{\rm Mgb\cdot(0.72\cdot Fe5270 + 0.28\cdot Fe5335)}$, which is independent of $\alpha$-enhancement and can be an effective indicator of stellar metallicity \citep{2003A&A...401..429T}. Figure \ref{fig:stellar_properties}(a) displays the {\hb} indices as functions of $\rm [MgFe]^{\prime}$ indices along the major axis, with the primary and secondary disks in red and blue, respectively. The model-grid is the prediction from single stellar population models given by \cite{2011MNRAS.412.2183T}. Comparing the data points with the model-grid, the secondary disk has younger stellar population and poorer stellar metallicity than the primary disk.

Figure \ref{fig:stellar_properties}(b) displays the D$_{n}$4000 indices as functions of radii along the major axis. The circles represent the results from the pPXF two kinematic component fit, with the primary and secondary disks in red and blue, respectively. The stellar population of the primary disk is old, with D$_{n}$4000 of $\sim$1.8. Meanwhile, the stellar population of the secondary disk is young in the disk region, presenting a constant D$_{n}$4000 of $\sim$1.4. The D$_{n}$4000 of the secondary disk increases in the central region, where the emission is bulge-dominated.

Figure \ref{fig:stellar_properties}(c) displays the stellar metallicity as functions of radii along the major axis, which is estimated by comparing the position of each circle with the model-grid in Figure \ref{fig:stellar_properties}(a). The stellar metallicity of the primary disk is rich, and follows a constant distribution. Meanwhile, the stellar metallicity of the secondary disk is poor, and follows a negative gradient with metallicity decreasing on the outskirts. Given the bulge-dominated feature in the central region displayed in Figure \ref{fig:stellar_properties}(b), we don't overinterpret the negative gradient here. Moreover, we estimate the stellar population age for two stellar disks in Figure \ref{fig:stellar_properties}(a), and find the stellar population of the primary disk being older than that of the secondary disk, which is consistent with the D$_{n}$4000 indices in Figure \ref{fig:stellar_properties}(b).

\section{Formation of counter-rotating stellar disks in SDSS~J0748+4441}
\label{sec: discussion}

% Internal and external processes
Early studies proposed that internal structures such as bulge and bar \citep{1994ApJ...420L..67E} are responsible for the formation of CRDs. However, these scenarios failed to explain the presence of different stellar populations of two stellar disks \citep{2013A&A...549A...3C}, which instead can be the results of external processes such as mergers and gas accretion \citep{2014ASPC..486...51C}. Major merger is known to be a destructive event that heats the system significantly, destroys the morphology of galaxies, and results in the formation of an elliptical galaxy. On the other hand, gas-rich minor merger and gas accretion are milder processes delivering the external gas that is counter-rotating with the pre-existing gas, and have weaker influence on the galaxy morphology \citep{2005A&A...437...69B}.

% Simulation: merger and gas accretion
With the development of long-slit spectroscopic and IFS instruments, studies on both individual galaxies \citep{2011MNRAS.412L.113C, 2013MNRAS.428.1296J} and galaxy samples \citep{2022MNRAS.511..139B, 2022ApJ...926L..13B} confirmed that the external processes including mergers and gas accretion contribute to the formation of CRDs. The common point in these processes is that the lately forming stellar disk inherits the angular momentum from the external gas, which is counter-rotating with the pre-existing stellar disk. During the formation of CRDs, once the external gas with opposite direction is abundant enough to dominate the gas rotation, the misalignment between gas and stars will be observed in galactic scale \citep{2016NatCo...713269C, 2022MNRAS.511.4685X}. \cite{2011MNRAS.414..968D} found that the gas and stars are misaligned in $\sim$42~percent early type galaxies in ATLAS$^{\rm 3D}$. However, assuming mergers as the only source of misalignment in simulation, \cite{2015MNRAS.448.1271L} obtained only $\sim$2-5~percent early type galaxies with misaligned gas and stars. After adding gas accretion in their simulation, the misalignment between gas and stars was found in $\sim$46~percent early type galaxies. These studies proved that gas accretion is a key process of bringing the external gas with opposite direction.

\begin{figure*}
\resizebox{1\textwidth}{!}{\includegraphics{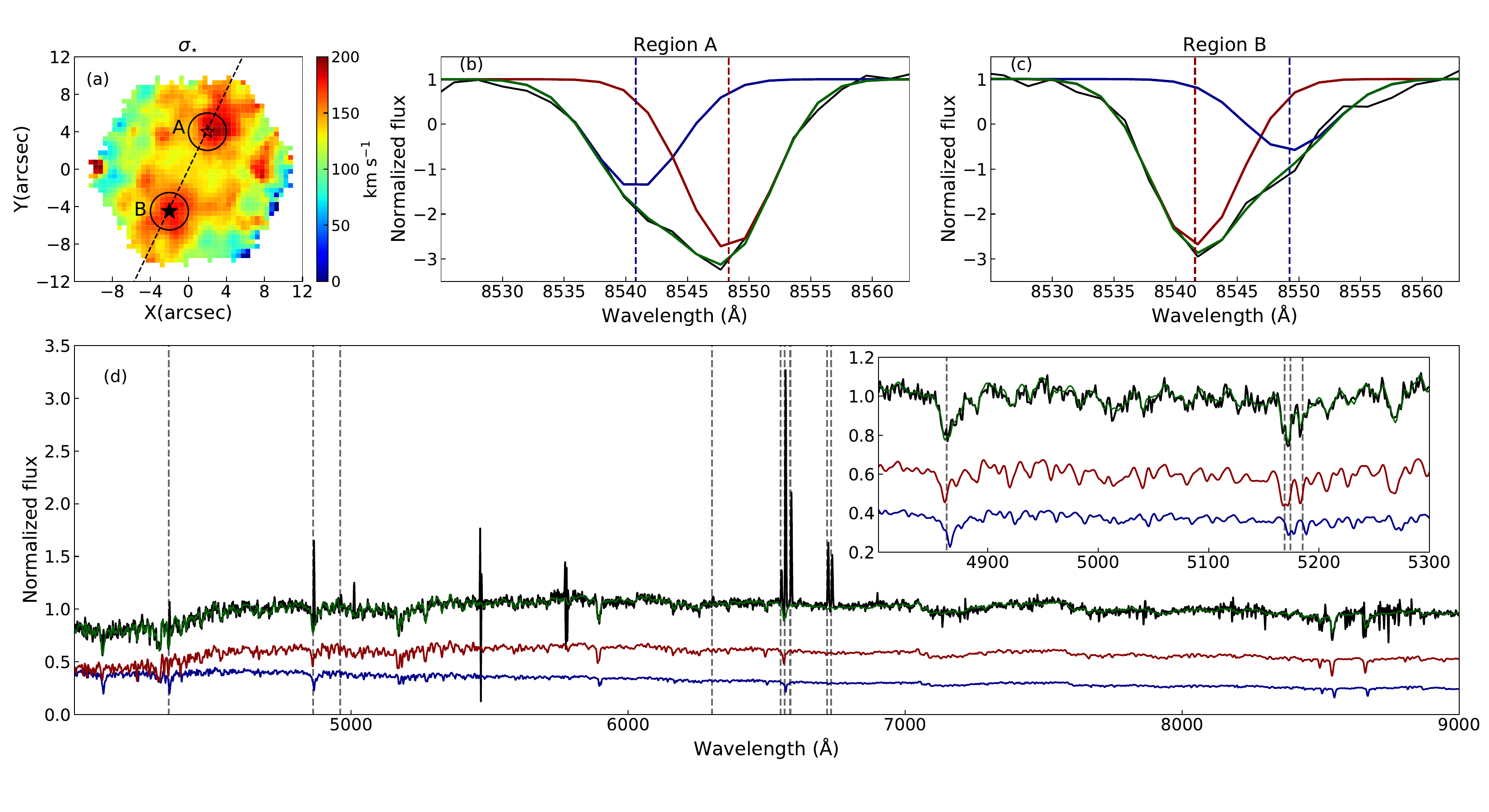}}
\caption{Spectral decomposition. (a) Stellar velocity dispersion field with median spectral SNR per spaxel higher than 3. The black empty and filled stars mark the 2$\sigma$ peaks. Two black rings outline the regions around the 2$\sigma$ peaks, with the top one named `Region A' and the bottom one named `Region B'. (b) \& (c) The black profiles show the stacked {\ca}~$\lambda$8542 lines within Regions A and B, respectively. The red and blue profiles show the double Gaussian fit on the {\ca}~$\lambda$8542 lines, with the stronger absorption in red and weaker absorption in blue. The red and blue dashed lines mark the line centers for corresponding models. The green profiles show the best-fit models as a combination of red and blue components. (d) The black profile shows the observed spectrum extracted from the filled star in panel (a). The grey dashed lines mark the rest-frame wavelength of emission lines that are masked prior to fit. The red and blue profiles show the two optimal models obtained by the pPXF two kinematic component fit, with primary and secondary disks in red and blue, respectively. The green profile is the best-fit model as a combination of two models. The insert drawing highlights the spectrum with 4800-5300~\AA~wavelength range. The four grey dashed lines mark the rest-frame wavelength of {\hb} and {\mg} triplet lines.}
\label{fig:spectrum_fit}
\end{figure*}

\begin{figure*}
    \resizebox{0.98\textwidth}{!}{\includegraphics{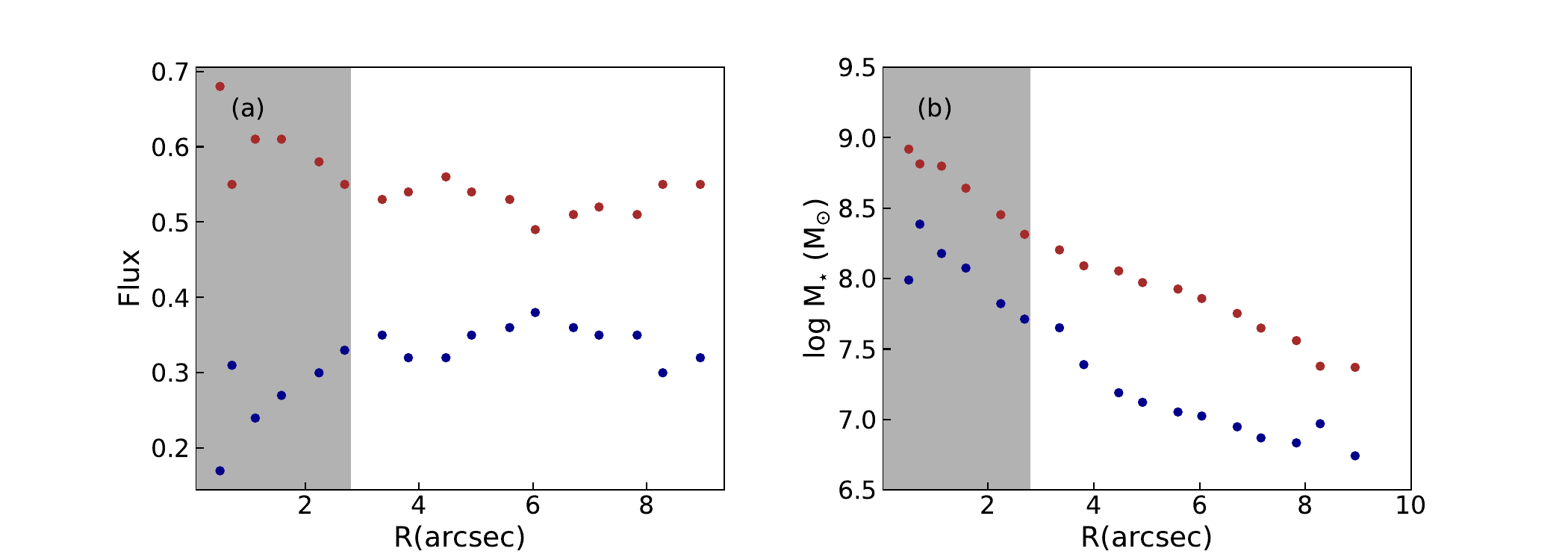}}
    \caption{Contributions of two stellar disks along the major axis. (a) \& (b) The circles show the fluxes and stellar masses as functions of radii, with the primary and secondary disks in red and blue.}
    \label{fig:disks_ratio}
\end{figure*}

% Flux and mass ratio between two disks
Based on the pPXF two kinematic component fit as displayed in Figure \ref{fig:spectrum_fit}(d), we calculate the flux contributions of two stellar disks in each spaxel. Figure \ref{fig:disks_ratio}(a) displays the flux contributions as functions of radii along the major axis, with the primary and secondary disks in red and blue, respectively. The primary disk is more luminous than the secondary disk across the whole galaxy. The average flux ratio between the primary and secondary disks is $\sim$1.8. We also estimate the masses of two stellar disks based on the decomposed spectra. For each disk model, we convolve the model spectrum with the SDSS $u$, $g$, $r$, $i$, $z$ filters to get the AB magnitudes. The stellar mass-to-light ratio is then obtained by comparing the $u$, $g$, $r$, $i$, $z$ magnitudes of each stellar disk with mimic galaxies generated in \cite{2012MNRAS.421..314C}. We refer readers to \cite{2012MNRAS.421..314C} for more details of the methodology. Figure \ref{fig:disks_ratio}(b) displays the stellar masses of two stellar disks as functions of radii, with the primary and secondary disks in red and blue, respectively. The primary disk is more massive than the secondary disk across the whole galaxy, with mass ratio of $\sim$5.2.

\begin{figure*}
\resizebox{0.98\textwidth}{!}{\includegraphics{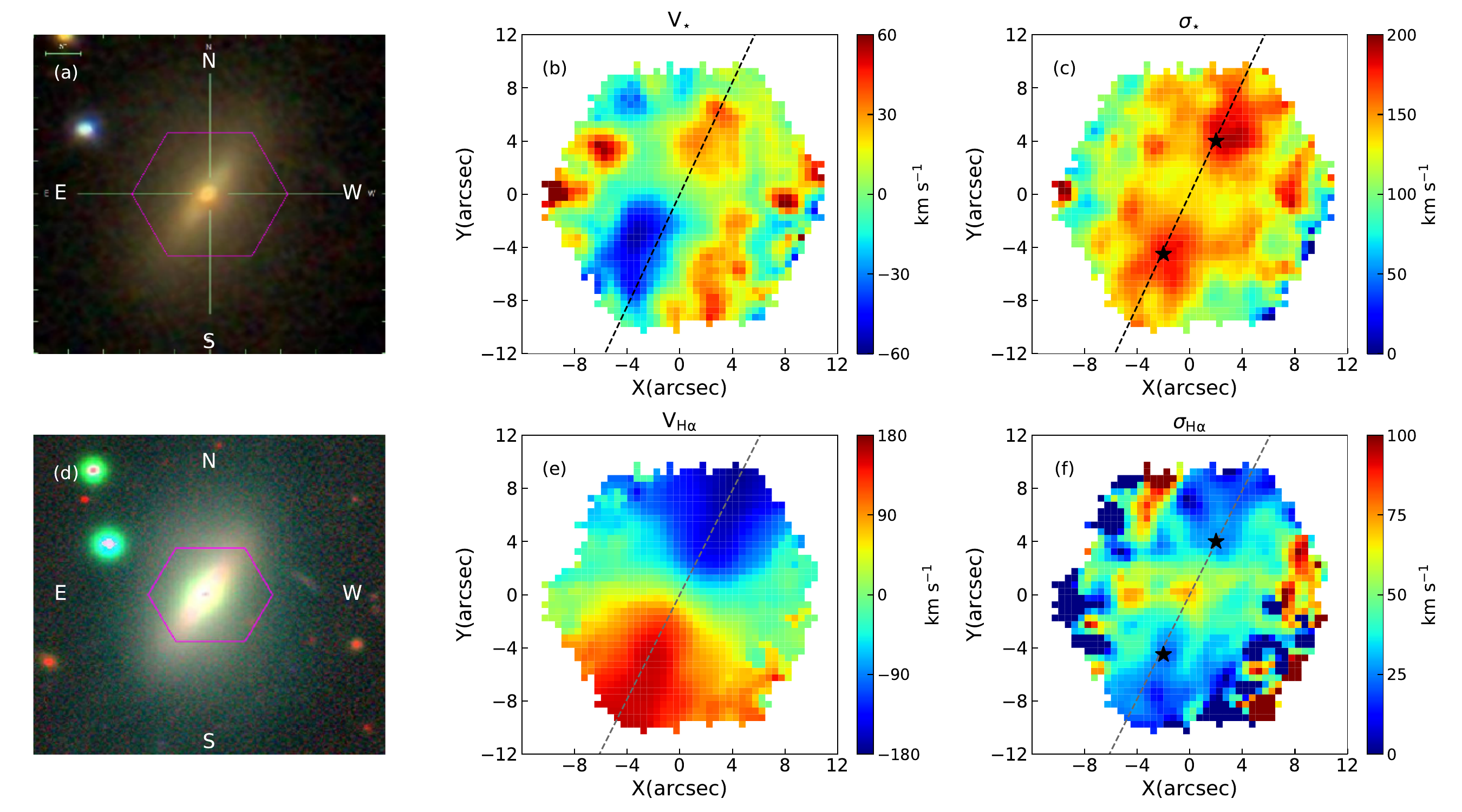}}
\caption{Images and spatially resolved kinematics. (a) SDSS $g$, $r$, $i$ color image. (b) \& (c) Stellar velocity and velocity dispersion fields for spaxels with median spectral SNR per spaxel higher than 3. The black dashed lines show the photometric major axis in Figures \ref{fig:kinematics} to \ref{fig:spectrum_fit}. The black stars mark the 2$\sigma$ peaks along the major axis in Figures \ref{fig:kinematics} and \ref{fig:emission}. (d) DESI $g$, $r$, $z$ color image. (e) \& (f) Gas velocity and velocity dispersion fields for spaxels with {\ha} emission line SNR higher than 3. The grey dashed line shows the kinematic major axis.}
\label{fig:kinematics}
\end{figure*}

\begin{figure*}
    \resizebox{0.98\textwidth}{!}{\includegraphics{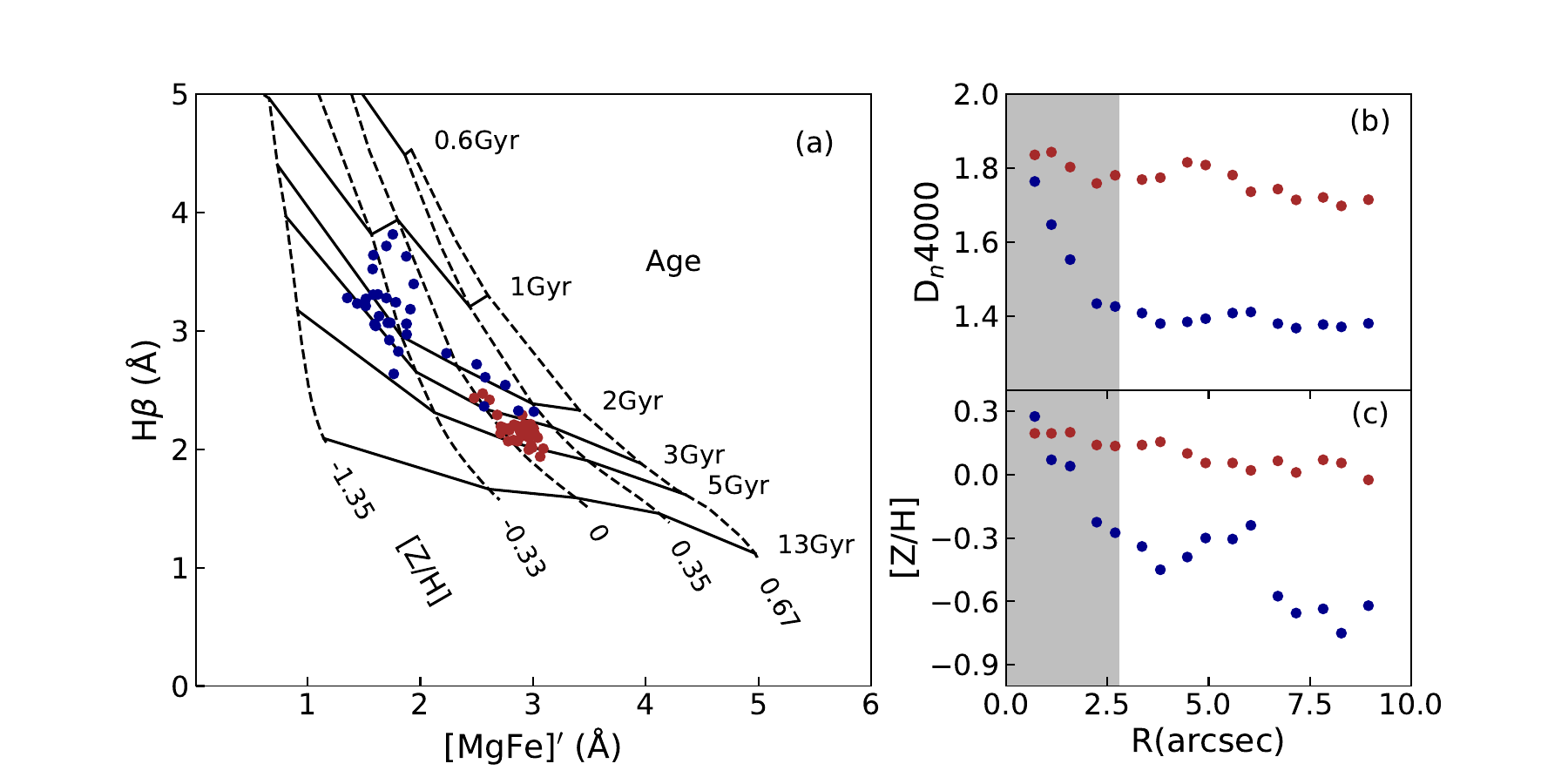}}
    \caption{Properties of two stellar disks along the major axis. (a) The circles show H$\beta$ indices as functions of $\rm [MgFe]^{\prime}$ indices, with the primary and secondary disks in red and blue. The model-grid is the prediction from single stellar population models given by \citet{2011MNRAS.412.2183T}. (b) \& (c) The circles show the D$_{n}$4000 indices and stellar metallicity as functions of radii, with the primary and secondary disks in red and blue.}
    \label{fig:stellar_properties}
\end{figure*}

% Formation scenario: gas-rich minor merger
% No interaction remnant & stellar population age of secondary disk ~ 2 Gyr
A gas-rich minor merger in retrograde orbit can provide abundant gas that is counter-rotating with the pre-existing stars, and fuel the formation of secondary disk. \cite{2014A&A...566A..97J} simulated mergers with different mass ratios, and found that the merger remnant features can be observed for $\sim$5.8~Gyrs at $r$-band magnitude limit of $\sim$25~mag~arcsec$^{-2}$ for a merger with a mass ratio of 6. The stellar population age of the secondary disk in SDSS~J0748+4441 is $\sim$2~Gyrs in Figure \ref{fig:stellar_properties}(b), suggesting that the gas-rich minor merger should happen $\sim$2~Gyrs ago. On the other hand, the DESI $g$, $r$, $z$ color image of SDSS~J0748+4441 is displayed in Figure \ref{fig:kinematics}(d), which is much deeper than 25~mag~arcsec$^{-2}$ in $r$-band. We follow the method described in \cite{2021MNRAS.501...14L} to check whether there is any merger remnant feature, finding that it is an isolated galaxy. The gas-rich minor merger is inapplicable to explain the formation of CRDs in SDSS~J0748+4441 due to the lack of merger remnant features.

% Formation scenario: gas accretion
We propose gas accretion in retrograde orbit as formation scenario for CRDs in SDSS~J0748+4441. Given the stellar population age of the secondary disk in Figure \ref{fig:stellar_properties}(b), the gas accretion in this galaxy happened $\sim$2~Gyrs ago. The global stellar mass of this galaxy in the MaNGA DRP is $\sim$10$^{10}$~M$_{\odot}$. Combining the mass ratio of $\sim$5.2 between two stellar disks (Figure \ref{fig:disks_ratio}b), accretion brought at least 1.6$\times$10$^{9}$~M$_{\odot}$ counter-rotating gas into SDSS~J0748+4441. The gas accretion can originate from galaxy environment including the cosmic web \citep{2016MNRAS.461.2068K} or a gas-rich companion \citep{2011MNRAS.412L.113C}.
\\ \hspace*{\fill} \\
{\noindent \bf Acknowledgements}
We are grateful to Can Xu for her valuable comments. M. B. acknowledges support from the National Natural Science Foundation of China, NSFC Grant No. 12303009. Y. C. acknowledges support from the National Natural Science Foundation of China, NSFC Grant No. 12333002, 11573013, 11733002, 11922302, the China Manned Space Project with NO. CMS-CSST-2021-A05. M. Y. acknowledges support from the China Postdoctoral Science Foundation, Grant No. 2021M703337. This work was supported by the research grants from the China Manned Space Project, the second-stage CSST science project `Investigation of small-scale structures in galaxies and forecasting of observations'.

Funding for the Sloan Digital Sky Survey IV has been provided by the Alfred P. Sloan Foundation, the U.S. Department of Energy Office of Science, and the Participating Institutions. SDSS- IV acknowledges support and resources from the Center for High-Performance Computing at the University of Utah. The SDSS web site is www.sdss.org. SDSS-IV is managed by the Astrophysical Research Consortium for the Participating Institutions of the SDSS Collaboration including the Brazilian Participation Group, the Carnegie Institution for Science, Carnegie Mellon University, the Chilean Participation Group, the French Participation Group, Harvard-Smithsonian Center for Astrophysics, Instituto de Astrof\'{i}sica de Canarias, The Johns Hopkins University, Kavli Institute for the Physics and Mathematics of the Universe (IPMU) / University of Tokyo, Lawrence Berkeley National Laboratory, Leibniz Institut  f\"{u}r Astrophysik Potsdam (AIP), Max-Planck-Institut  f\"{u}r   Astronomie  (MPIA  Heidelberg), Max-Planck-Institut   f\"{u}r   Astrophysik  (MPA   Garching), Max-Planck-Institut f\"{u}r Extraterrestrische Physik (MPE), National Astronomical Observatory of China, New Mexico State University, New York University, University of Notre Dame, Observat\'{o}rio Nacional / MCTI, The Ohio State University, Pennsylvania State University, Shanghai Astronomical Observatory, United Kingdom Participation Group, Universidad Nacional  Aut\'{o}noma de M\'{e}xico,  University of Arizona, University of Colorado  Boulder, University of Oxford, University of Portsmouth, University of Utah, University of Virginia, University  of Washington,  University of  Wisconsin, Vanderbilt University, and Yale University.
\\ \hspace*{\fill} \\
{\noindent \bf Data Availability}
The data underlying this article will be shared on reasonable request to the corresponding author.

\bsp

\label{lastpage}

\begin{thebibliography}{}

    \bibitem[\protect\citeauthoryear{Baldwin, Phillips, \& Terlevich}{1981}]{1981PASP...93....5B} Baldwin J.~A., Phillips M.~M., Terlevich R., 1981, PASP, 93, 5. doi:10.1086/130766

    \bibitem[\protect\citeauthoryear{Bao et al.}{2022}]{2022ApJ...926L..13B} Bao M., Chen Y., Zhu P., Shi Y., Bizyaev D., Zhu L., Yang M., et al., 2022, ApJL, 926, L13. doi:10.3847/2041-8213/ac52ad

    \bibitem[\protect\citeauthoryear{Bevacqua, Cappellari, \& Pellegrini}{2022}]{2022MNRAS.511..139B} Bevacqua D., Cappellari M., Pellegrini S., 2022, MNRAS, 511, 139. doi:10.1093/mnras/stab3732

    \bibitem[\protect\citeauthoryear{Blanton et al.}{2017}]{2017AJ....154...28B} Blanton M.~R., Bershady M.~A., Abolfathi B., Albareti F.~D., Allende Prieto C., Almeida A., Alonso-Garc{\'\i}a J., et al., 2017, AJ, 154, 28. doi:10.3847/1538-3881/aa7567

    \bibitem[\protect\citeauthoryear{Bournaud, Jog, \& Combes}{2005}]{2005A&A...437...69B} Bournaud F., Jog C.~J., Combes F., 2005, A\&A, 437, 69. doi:10.1051/0004-6361:20042036

    \bibitem[\protect\citeauthoryear{Bundy et al.}{2015}]{2015ApJ...798....7B} Bundy K., Bershady M.~A., Law D.~R., Yan R., Drory N., MacDonald N., Wake D.~A., et al., 2015, ApJ, 798, 7. doi:10.1088/0004-637X/798/1/7
    
    \bibitem[\protect\citeauthoryear{Cappellari}{2002}]{2002MNRAS.333..400C} Cappellari M., 2002, MNRAS, 333, 400. doi:10.1046/j.1365-8711.2002.05412.x

    \bibitem[\protect\citeauthoryear{Cappellari}{2017}]{2017MNRAS.466..798C} Cappellari M., 2017, MNRAS, 466, 798. doi:10.1093/mnras/stw3020

    \bibitem[\protect\citeauthoryear{Chen et al.}{2012}]{2012MNRAS.421..314C} Chen Y.-M., Kauffmann G., Tremonti C.~A., White S., Heckman T.~M., Kova{\v{c}} K., Bundy K., et al., 2012, MNRAS, 421, 314. doi:10.1111/j.1365-2966.2011.20306.x

    \bibitem[\protect\citeauthoryear{Chen et al.}{2016}]{2016NatCo...713269C} Chen Y.-M., Shi Y., Tremonti C.~A., Bershady M., Merrifield M., Emsellem E., Jin Y.-F., et al., 2016, NatCo, 7, 13269. doi:10.1038/ncomms13269

    \bibitem[\protect\citeauthoryear{Coccato et al.}{2011}]{2011MNRAS.412L.113C} Coccato L., Morelli L., Corsini E.~M., Buson L., Pizzella A., Vergani D., Bertola F., 2011, MNRAS, 412, L113. doi:10.1111/j.1745-3933.2011.01016.x

    \bibitem[\protect\citeauthoryear{Coccato et al.}{2013}]{2013A&A...549A...3C} Coccato L., Morelli L., Pizzella A., Corsini E.~M., Buson L.~M., Dalla Bont{\`a} E., 2013, A\&A, 549, A3. doi:10.1051/0004-6361/201220460

    \bibitem[\protect\citeauthoryear{Corsini}{2014}]{2014ASPC..486...51C} Corsini E.~M., 2014, ASPC, 486, 51. doi:10.48550/arXiv.1403.1263

    \bibitem[\protect\citeauthoryear{Davis et al.}{2011}]{2011MNRAS.414..968D} Davis T.~A., Bureau M., Young L.~M., Alatalo K., Blitz L., Cappellari M., Scott N., et al., 2011, MNRAS, 414, 968. doi:10.1111/j.1365-2966.2011.18284.x

    \bibitem[\protect\citeauthoryear{Ding et al.}{2023}]{2023A&A...672A..84D} Ding Y., Zhu L., van de Ven G., Coccato L., Corsini E.~M., Costantin L., Fahrion K., et al., 2023, A\&A, 672, A84. doi:10.1051/0004-6361/202244558

    \bibitem[\protect\citeauthoryear{Emsellem, Monnet, \& Bacon}{1994}]{1994A&A...285..723E} Emsellem E., Monnet G., Bacon R., 1994, A\&A, 285, 723

    \bibitem[\protect\citeauthoryear{Evans \& Collett}{1994}]{1994ApJ...420L..67E} Evans N.~W., Collett J.~L., 1994, ApJL, 420, L67. doi:10.1086/187164

    \bibitem[\protect\citeauthoryear{Falc{\'o}n-Barroso et al.}{2011}]{2011A&A...532A..95F} Falc{\'o}n-Barroso J., S{\'a}nchez-Bl{\'a}zquez P., Vazdekis A., Ricciardelli E., Cardiel N., Cenarro A.~J., Gorgas J., et al., 2011, A\&A, 532, A95. doi:10.1051/0004-6361/201116842

    
    \bibitem[\protect\citeauthoryear{Falc{\'o}n-Barroso \& Martig}{2021}]{2021A&A...646A..31F} Falc{\'o}n-Barroso J., Martig M., 2021, A\&A, 646, A31. doi:10.1051/0004-6361/202039624

    \bibitem[\protect\citeauthoryear{Gunn et al.}{2006}]{2006AJ....131.2332G} Gunn J.~E., Siegmund W.~A., Mannery E.~J., Owen R.~E., Hull C.~L., Leger R.~F., Carey L.~N., et al., 2006, AJ, 131, 2332. doi:10.1086/500975

    \bibitem[\protect\citeauthoryear{Jethwa et al.}{2020}]{2020ascl.soft11007J} Jethwa P., Thater S., Maindl T., Van de Ven G., 2020, ascl.soft. ascl:2011.007

    \bibitem[\protect\citeauthoryear{Ji, Peirani, \& Yi}{2014}]{2014A&A...566A..97J} Ji I., Peirani S., Yi S.~K., 2014, A\&A, 566, A97. doi:10.1051/0004-6361/201423530

    \bibitem[\protect\citeauthoryear{Jin et al.}{2020}]{2020MNRAS.491.1690J} Jin Y., Zhu L., Long R.~J., Mao S., Wang L., van de Ven G., 2020, MNRAS, 491, 1690. doi:10.1093/mnras/stz3072

    \bibitem[\protect\citeauthoryear{Johnston et al.}{2013}]{2013MNRAS.428.1296J} Johnston E.~J., Merrifield M.~R., Arag{\'o}n-Salamanca A., Cappellari M., 2013, MNRAS, 428, 1296. doi:10.1093/mnras/sts121

    \bibitem[\protect\citeauthoryear{Kaiser et al.}{2002}]{2002SPIE.4836..154K} Kaiser N., Aussel H., Burke B.~E., Boesgaard H., Chambers K., Chun M.~R., Heasley J.~N., et al., 2002, SPIE, 4836, 154. doi:10.1117/12.457365

    \bibitem[\protect\citeauthoryear{Katkov et al.}{2011}]{2011BaltA..20..453K} Katkov I., Chilingarian I., Sil'chenko O., Zasov A., Afanasiev V., 2011, BaltA, 20, 453. doi:10.1515/astro-2017-0318

    \bibitem[\protect\citeauthoryear{Katkov et al.}{2016}]{2016MNRAS.461.2068K} Katkov I.~Y., Sil'chenko O.~K., Chilingarian I.~V., Uklein R.~I., Egorov O.~V., 2016, MNRAS, 461, 2068. doi:10.1093/mnras/stw1452

    \bibitem[\protect\citeauthoryear{Kauffmann et al.}{2003}]{2003MNRAS.346.1055K} Kauffmann G., Heckman T.~M., Tremonti C., Brinchmann J., Charlot S., White S.~D.~M., Ridgway S.~E., et al., 2003, MNRAS, 346, 1055. doi:10.1111/j.1365-2966.2003.07154.x

    \bibitem[\protect\citeauthoryear{Kewley et al.}{2001}]{2001ApJ...556..121K} Kewley L.~J., Dopita M.~A., Sutherland R.~S., Heisler C.~A., Trevena J., 2001, ApJ, 556, 121. doi:10.1086/321545

    \bibitem[\protect\citeauthoryear{Krajnovi{\'c} et al.}{2006}]{2006MNRAS.366..787K} Krajnovi{\'c} D., Cappellari M., de Zeeuw P.~T., Copin Y., 2006, MNRAS, 366, 787. doi:10.1111/j.1365-2966.2005.09902.x

    \bibitem[\protect\citeauthoryear{Krajnovi{\'c} et al.}{2011}]{2011MNRAS.414.2923K} Krajnovi{\'c} D., Emsellem E., Cappellari M., Alatalo K., Blitz L., Bois M., Bournaud F., et al., 2011, MNRAS, 414, 2923. doi:10.1111/j.1365-2966.2011.18560.x

    \bibitem[\protect\citeauthoryear{Lagos et al.}{2015}]{2015MNRAS.448.1271L} Lagos C. del P., Padilla N.~D., Davis T.~A., Lacey C.~G., Baugh C.~M., Gonzalez-Perez V., Zwaan M.~A., et al., 2015, MNRAS, 448, 1271. doi:10.1093/mnras/stu2763

    \bibitem[\protect\citeauthoryear{Law et al.}{2016}]{2016AJ....152...83L} Law D.~R., Cherinka B., Yan R., Andrews B.~H., Bershady M.~A., Bizyaev D., Blanc G.~A., et al., 2016, AJ, 152, 83. doi:10.3847/0004-6256/152/4/83

    \bibitem[\protect\citeauthoryear{Li et al.}{2021}]{2021MNRAS.501...14L} Li S.-. lin ., Shi Y., Bizyaev D., Duckworth C., Yan R.-. bin ., Chen Y.-. mei ., Bing L.-. ji ., et al., 2021, MNRAS, 501, 14. doi:10.1093/mnras/staa3618

    \bibitem[\protect\citeauthoryear{Mitzkus, Cappellari, \& Walcher}{2017}]{2017MNRAS.464.4789M} Mitzkus M., Cappellari M., Walcher C.~J., 2017, MNRAS, 464, 4789. doi:10.1093/mnras/stw2677

    \bibitem[\protect\citeauthoryear{Rubino et al.}{2021}]{2021A&A...654A..30R} Rubino M., Pizzella A., Morelli L., Coccato L., Portaluri E., Debattista V.~P., Corsini E.~M., et al., 2021, A\&A, 654, A30. doi:10.1051/0004-6361/202140702


    \bibitem[\protect\citeauthoryear{Pizzella et al.}{2014}]{2014A&A...570A..79P} Pizzella A., Morelli L., Corsini E.~M., Dalla Bont{\`a} E., Coccato L., Sanjana G., 2014, A\&A, 570, A79. doi:10.1051/0004-6361/201424746

    \bibitem[\protect\citeauthoryear{Pizzella et al.}{2018}]{2018A&A...616A..22P} Pizzella A., Morelli L., Coccato L., Corsini E.~M., Dalla Bont{\`a} E., Fabricius M., Saglia R.~P., 2018, A\&A, 616, A22. doi:10.1051/0004-6361/201731712

    \bibitem[\protect\citeauthoryear{Rix et al.}{1992}]{1992ApJ...400L...5R} Rix H.-W., Franx M., Fisher D., Illingworth G., 1992, ApJL, 400, L5. doi:10.1086/186635

    \bibitem[\protect\citeauthoryear{Rubin, Graham, \& Kenney}{1992}]{1992ApJ...394L...9R} Rubin V.~C., Graham J.~A., Kenney J.~D.~P., 1992, ApJL, 394, L9. doi:10.1086/186460
    
    \bibitem[\protect\citeauthoryear{S{\'a}nchez-Bl{\'a}zquez et al.}{2006}]{2006MNRAS.371..703S} S{\'a}nchez-Bl{\'a}zquez P., Peletier R.~F., Jim{\'e}nez-Vicente J., Cardiel N., Cenarro A.~J., Falc{\'o}n-Barroso J., Gorgas J., et al., 2006, MNRAS, 371, 703. doi:10.1111/j.1365-2966.2006.10699.x

    \bibitem[\protect\citeauthoryear{S{\'a}nchez et al.}{2016}]{2016RMxAA..52...21S} S{\'a}nchez S.~F., P{\'e}rez E., S{\'a}nchez-Bl{\'a}zquez P., Gonz{\'a}lez J.~J., Ros{\'a}les-Ortega F.~F., Cano-D{\'\i}az M., L{\'o}pez-Cob{\'a} C., et al., 2016, RMxAA, 52, 21. doi:10.48550/arXiv.1509.08552

    \bibitem[\protect\citeauthoryear{S{\'a}nchez et al.}{2018}]{2018RMxAA..54..217S} S{\'a}nchez S.~F., Avila-Reese V., Hernandez-Toledo H., Cortes-Su{\'a}rez E., Rodr{\'\i}guez-Puebla A., Ibarra-Medel H., Cano-D{\'\i}az M., et al., 2018, RMxAA, 54, 217. doi:10.48550/arXiv.1709.05438

    \bibitem[\protect\citeauthoryear{Santucci et al.}{2022}]{2022ApJ...930..153S} Santucci G., Brough S., van de Sande J., McDermid R.~M., van de Ven G., Zhu L., D'Eugenio F., et al., 2022, ApJ, 930, 153. doi:10.3847/1538-4357/ac5bd5

    \bibitem[\protect\citeauthoryear{Schwarzschild}{1979}]{1979ApJ...232..236S} Schwarzschild M., 1979, ApJ, 232, 236. doi:10.1086/157282

    \bibitem[\protect\citeauthoryear{Smee et al.}{2013}]{2013AJ....146...32S} Smee S.~A., Gunn J.~E., Uomoto A., Roe N., Schlegel D., Rockosi C.~M., Carr M.~A., et al., 2013, AJ, 146, 32. doi:10.1088/0004-6256/146/2/32

    \bibitem[\protect\citeauthoryear{Tahmasebzadeh et al.}{2022}]{2022ApJ...941..109T} Tahmasebzadeh B., Zhu L., Shen J., Gerhard O., van de Ven G., 2022, ApJ, 941, 109. doi:10.3847/1538-4357/ac9df6

    \bibitem[\protect\citeauthoryear{Thater et al.}{2022}]{2022A&A...667A..51T} Thater S., Jethwa P., Tahmasebzadeh B., Zhu L., den Brok M., Santucci G., Ding Y., et al., 2022, A\&A, 667, A51. doi:10.1051/0004-6361/202243926

    \bibitem[\protect\citeauthoryear{Thomas, Maraston, \& Johansson}{2011}]{2011MNRAS.412.2183T} Thomas D., Maraston C., Johansson J., 2011, MNRAS, 412, 2183. doi:10.1111/j.1365-2966.2010.18049.x
    
    \bibitem[\protect\citeauthoryear{Thomas \& Maraston}{2003}]{2003A&A...401..429T} Thomas D., Maraston C., 2003, A\&A, 401, 429. doi:10.1051/0004-6361:20030153

    \bibitem[\protect\citeauthoryear{van den Bosch et al.}{2008}]{2008MNRAS.385..647V} van den Bosch R.~C.~E., van de Ven G., Verolme E.~K., Cappellari M., de Zeeuw P.~T., 2008, MNRAS, 385, 647. doi:10.1111/j.1365-2966.2008.12874.x

    \bibitem[\protect\citeauthoryear{Westfall et al.}{2019}]{2019AJ....158..231W} Westfall K.~B., Cappellari M., Bershady M.~A., Bundy K., Belfiore F., Ji X., Law D.~R., et al., 2019, AJ, 158, 231. doi:10.3847/1538-3881/ab44a2

    \bibitem[\protect\citeauthoryear{Worthey et al.}{1994}]{1994ApJS...94..687W} Worthey G., Faber S.~M., Gonzalez J.~J., Burstein D., 1994, ApJS, 94, 687. doi:10.1086/192087
    
    \bibitem[\protect\citeauthoryear{Xu et al.}{2022}]{2022MNRAS.511.4685X} Xu H., Chen Y., Shi Y., Zhou Y., Bizyaev D., Bao M., Beom M., et al., 2022, MNRAS, 511, 4685. doi:10.1093/mnras/stac354

    \bibitem[\protect\citeauthoryear{Zhu et al.}{2018}]{2018MNRAS.479..945Z} Zhu L., van de Ven G., M{\'e}ndez-Abreu J., Obreja A., 2018, MNRAS, 479, 945. doi:10.1093/mnras/sty1521

    \bibitem[\protect\citeauthoryear{Zhu et al.}{2018}]{2018MNRAS.473.3000Z} Zhu L., van den Bosch R., van de Ven G., Lyubenova M., Falc{\'o}n-Barroso J., Meidt S.~E., Martig M., et al., 2018, MNRAS, 473, 3000. doi:10.1093/mnras/stx2409

\end{thebibliography}
\end{document}